\documentclass[useAMS,usenatbib]{mn2e}
\usepackage{amsmath}
\usepackage{blindtext}
\usepackage{graphicx}
\usepackage{amssymb}
\usepackage{multirow}
\usepackage{verbatim}   
\usepackage{color}     
\usepackage{subfigure} 
\usepackage{hyperref}
\usepackage{latexsym}
\usepackage{caption}
\usepackage{wrapfig}
\usepackage{soul}
\usepackage{natbib}
\usepackage{float} 
\usepackage{placeins}

\title[Testing bound dark energy with cosmological parameter and fundamental constant evolution]{Testing bound dark energy with cosmological parameter and fundamental constant evolution}
\author[Rodger I. Thompson]{Rodger I. Thompson$^{1}$\thanks{E-mail:
rit@email.arizona.edu (RIT)}\\
$^{1}$Steward Observatory, University of Arizona, Tucson, AZ 85721, USA}

\begin{document}

\date{Accepted xxxx. Received xxxx; in original form xxxx}

\pagerange{\pageref{firstpage}--\pageref{lastpage}} \pubyear{2016}

\maketitle

\label{firstpage}

\begin{abstract}
A new bound dark energy, BDE, cosmology has been proposed where the dark energy is the binding
energy between light meson fields that condense a few tens of years after the big bang.  It is reported
that the correct dark energy density emerges using particle physics without fine tuning.  This alone
makes the BDE cosmology worthy of further investigation. This work looks at the late time BDE
predictions of the evolution of cosmological parameters and the values of fundamental constants 
to determine whether the cosmology's predictions are consistent with observation.  The work
considers the time period between a scale factor of 0.1 and 1.0.  A model BDE cosmology is
considered with current day values of the cosmological parameters well within the observational
limits.  The calculations use three different values of the current day dark energy equation of state
close to minus one.  All three cases produce evolutions of the cosmological parameters and fundamental 
constants consistent with the observational constraints.  Analytic relations between the BDE and 
cosmological parameters are developed to insure a consistent set of parameters.
\end{abstract}


\maketitle

\section{Introduction} \label{s-intro} 
This paper examines the late time, scale factor $a$ = 0.1 to 1.0, evolution of cosmological parameters 
and fundamental constants in a Bound Dark Energy, BDE, cosmology recently introduced in
detail by \citet{alm19}, hereinafter AM19 and summarized by \citet{mac18}
hereinafter MA18.  As  described in AM19 and MA18, hereinafter AMMA, a very light 
meson field is postulated that is initially massless but condenses into massive scalars
at a condensation scale $\Lambda_c$ at a scale factor $a_c$.  The remarkable feature
of the BDE cosmology is that it predicts a dark energy density that is compatible with
the dark energy density required for the observed evolution of the universe including
late time inflation.  The goal of this work is to calculate the late time evolution of
cosmological parameters and fundamental constants using the BDE cosmology to see
if it is consistent with the current observational measurements and constraints.   Natural units 
are used and masses are given in reduced Planck mass  $M_p =\sqrt{\frac{\hbar c}{8 \pi G}}$ 
units to be consistent with general cosmological practice.  Note that AMMA use Planck masses
$m_p=\sqrt{\frac{\hbar c}{G}}$ consistent with general particle physics practice.

\section{Bound Dark Energy} \label{s-bde}
This is a  general description of bound dark energy to set the context for the following discussion.
Although the main purpose of this work is to test the theory by comparison with observables some
modifications of the AMMA results are provided.   AM19 describes Bound Dark Energy as follows.  
``BDE is derived from particle physics and corresponds to the lightest meson field $\phi$ dynamically 
formed at low energies due to the strong coupling constant.  The evolution of dark energy is determined
by the scalar potential $V(\phi)=\Lambda_c^{4+2/3}\phi^{-2/3}$ arising from non-perturbative 
effects at a condensation scale $\Lambda_c$ and a scale factor $a_c$.''  The BDE potential is a specific 
form of the general potential $V(\phi)=M^{4+p}\phi^{-p}$ with $p=2/3$.  The value of $p$ is determined 
by the parameters in the Affleck-Dine-Seiberg superpotential as discussed in AM19.  The units of $M$, 
$\Lambda_c$ and $\phi$ are mass expressed in terms of the reduced Planck mass $M_p$
\begin{equation} \label{eq-bdev}
V(\phi)=\Lambda_c^{4+2/3}\phi^{-2/3}
\end{equation}
It is called Bound Dark Energy because the dark energy is due to the binding energy of the meson fields.

The condensation scale and the condensation scale factor are related by (AMMA)
\begin{equation} \label{eq-ac}
a_c \Lambda_c = 1.0939 \times 10^{-4} eV = 4.49 \times 10^{-32} M_{p}
\end{equation}
Consistent with normal particle physics practices AMMA calculate the most likely current values of several 
cosmological parameters with a Markov-Chain Monte Carlo, MCMC, analysis using the CAMB and CosmoMC 
codes.  Table 1 of AM19 contains the results of the analysis. The MMC value of $\Lambda_c$ is $44.09 \pm 
0.28 eV$.  This is consistent with the theoretical value of $\Lambda_c = 34_{-11}^{16}eV$ except for the
much larger error limits which are mainly due to uncertainties in the QCD scale (AM19). This sets the value 
of the condensation scale factor at $2.48 \times 10^{-6}$, roughly 
67 years after the big bang.  Note that the value of $\phi$ at this time is $\Lambda_c$. This work takes a 
more cosmological approach using general cosmology tools such as the Einstein equations to explore the 
relationships between various parameters.  The goal is to also calculate accurate analytic functions of the 
fundamental constant and cosmological parameter evolution as functions of the scale factor $a$, not just 
the current parameter values.

The dark energy pressure and density in BDE are identical to the quintessence equations (AM19).
\begin{equation} \label{eq-rhop}
\rho_{\phi} \equiv \frac{\dot{\phi}^2}{2}+V(\phi), \hspace{1cm} p_{\phi}  \equiv \frac{\dot{\phi}^2}{2}-V(\phi)
\end{equation}
This means that much of the work done with the beta function methodology using quintessence relations
for the dark energy pressure and density \citep{bin15, cic17, thm18, thm19} is relevant to the BDE 
calculations.

\subsection{The Relationships Between Parameters} \label{ss-bder}
The goal of this section is to establish the relationships between the BDE and cosmological parameters
to determine appropriate constraints required by the relationships.  In considering the possible range
of $\Lambda_c$ determined by the cosmological parameters a constraint is set that $\Lambda_c$ must be 
within $2\sigma$ of the most likely value determined in AM19 using the $1\sigma$ values of the 
theoretical value rather than the much more restrictive MCMC value.  This is done through a set 
of derived relationships between the BDE and cosmological parameters. In particular the Hubble
parameter $H$, the condensation scale $\Lambda_c$, the ratio of the dark energy density to the
critical density $\Omega_{\phi}$, the scalar $\phi$ and the dark energy equation of state $w$ 
are inter-related. In the following the general form of the potential $\Lambda_c^{4+p}\phi^{-p}$
is used followed by the form with $p=\frac{2}{3}$. 

Since $V(\phi)$ is the dark energy density the first relationship is
\begin{equation} \label{eq-bphio}
 V(\phi)= \Lambda_c^{4+p}\phi^{-p} =3 \Omega_{\phi} (\frac{H}{\kappa})^2
\end{equation}
with $\kappa = \sqrt{\frac{8\pi G}{\hbar c}}= \frac{1}{M_p}$. Note that since this work uses 
masses expressed in units of the reduced Planck mass $\kappa$ has a value of one in these units.  
$\kappa$ is, however, retained in the equations to indicate the correct power mass units in the
equantions. Equation~\ref{eq-bphio} can be solved for $\phi$ to give
\begin{equation} \label{eq-phih}
(\kappa\phi)^2= \kappa^2 (\frac{\Lambda_c^{4+p}}{3\Omega_{\phi}(\frac{H}{\kappa})^2})^\frac{2}{p}, 
 \kappa^2 (\frac{\Lambda_c^{4+2/3}}{3\Omega_{\phi}(\frac{H}{\kappa})^2})^3
\end{equation}
where the comma separates the general solution from the BDE solution with $p=\frac{2}{3}$.
Note that $\kappa \phi$ is dimensionless.

The next step utilizes the relationship between the scalar $\phi$ and the dark energy equation of
state $w$ established by \cite{thm18} for inverse power law potentials.
\begin{equation} \label{eq-phio}
\kappa \phi = \frac{p}{\sqrt{3\Omega_{\phi}(w+1)}}
\end{equation}
where $\Omega_{\phi}$ is the ratio of the dark energy density to the critical density. Equation.~\ref{eq-phih} 
and eqn.~\ref{eq-phio} produce the  equation for $(w+1)$. 
\begin{equation} \label{eq-wo}
(w+1)=\frac{p^2 \Omega_{\phi}^{\frac{2}{p}-1}}{3\kappa^2}(\frac{3(\frac{H}{\kappa})^2}{\Lambda_c^{4+p}})^{\frac{2}{p}}, \frac{4 \Omega_{\phi}^2}{27\kappa^2}(\frac{3(\frac{H}{\kappa})^2}{\Lambda_c^{\frac{14}{3}}})^3
\end{equation}

Equations~\ref{eq-bphio} through~\ref{eq-wo} can be rearranged to give
\begin{equation} \label{eq-lamc}
\Lambda_c=[\frac{3p^p \Omega^{1-\frac{p}{2}} (\frac{H}{3\kappa})^2}{(3\kappa^2 (w+1))^{\frac{p}{2}}}]^{\frac{1}{p+4}}, [\frac{3 (\frac{4}{9})^{\frac{1}{3}} \Omega^{\frac{2}{3}} (\frac{H}{3\kappa})^2}{(3\kappa^2 (w+1))^{\frac{1}{3}}}]^{\frac{3}{14}}
\end{equation}
\begin{equation} \label{eq-h}
H=\frac{\kappa \Lambda_c^{2+\frac{p}{2}}}{\sqrt{3}}[\frac{3\kappa^2 (w+1)}{p^2\Omega_{\phi}^{\frac{2}{p}-1}}]^{\frac{p}{4}}, \frac{\kappa \Lambda_c^{\frac{7}{3}}}{\sqrt{3}}[\frac{27\kappa^2 (w+1)}{4\Omega_{\phi}^2}]^{\frac{1}{6}}
\end{equation}

\begin{figure}
\scalebox{.6}{\includegraphics{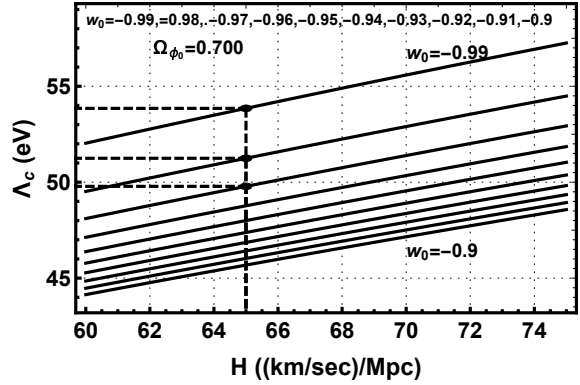}}
\caption{The condensation scale $\Lambda_c$ in eV is plotted versus the Hubble constant $H_0$ in 
(kilometer per second)/Mpc for $w_0$ values between -0.99 and -0.9 in 0.01 increments. The
black dots indicate the model $w_0$ choices (-0.99, -0.98. -0.97). The vertical dashed line indicates 
the chosen $H_0$ of 65 and the dashed horizontal lines indicate the corresponding values of $\Lambda_c$.} 
\label{fig-lch}
\end{figure}

The condensation scale does not evolve with time but its value is set by $H_0$, 
$w_0$ and $\Omega_{\phi_0}$ in eqn.~\ref{eq-lamc}.  The only cosmological parameter
that is allowed to have different initial values is the dark energy equation of state $w$.  
This requires a different $\Lambda_c$ value for each $w_0$ value as shown in fig~\ref{fig-lch}.
Figure~\ref{fig-lch} plots the condensation scale $\Lambda_c$ in eV versus the Hubble constant
$H_0$ for current values of the dark energy equation of state $w_0$ ranging from -0.99 to -0.9
in increments of 0.01.  The three $w_0$ values chosen in section~\ref{ss-mps} are shown as
black dots in the figure at $H_0=65$ with the horizontal dashed lines indicating the required
value of $\Lambda_c$ in eV for comparison with AM19.  Note that the greater the deviation of 
$w_0$ from minus one the lower the value of $\Lambda_c$ is.  Conversely the higher the value 
of $H_0$ is the higher the value of $\Lambda_c$ must be. 

The relationships developed in this section provide the means to calculate accurate evolutions of the
cosmological parameters and fundamental constants using the beta function methodology described in
section~\ref{s-bf} without numerical or MCMC calculations.

\section{The Beta Function Methodology} \label{s-bf}
The relationships developed in section~\ref{ss-bder} provide the tools to calculate accurate evolutions
as a function of the observable scale factor $a$ rather than of the unobservable scalar $\phi$ using the
beta function formalism \citep{bin15,cic17} as was done for a general set of potentials by \citep{thm18,thm19}. 
The beta function formalism is highly accurate when the current value of the dark energy equation of 
state $w_0$ is close to minus one. The beta function is defined as the derivative of the scalar $\phi$ 
with respect to the natural log of the scale factor $a$ \citep{bin15}
\begin{equation} \label{eq-beta}
\beta(\phi) \equiv \kappa\frac{d \phi}{d \ln(a)} = \kappa \phi'
\end{equation}
The prime on the right hand term denotes the derivative with respect to the natural 
log of the scale factor except when it denotes the integration variable inside an integral. It is 
clear that the beta function provides the link between the scalar $\phi$ and the scale factor $a$.

The beta function is not an arbitrary function of $\phi$ and $a$ but is determined by the 
model dark energy potential $V_m(\phi)$ such that \citep{cic17}
\begin{equation} \label{eq-mv}
V_m(\phi)=V_0\exp\{-\kappa\int \beta(\phi)d\phi\}
\end{equation}
where $V_m(\phi)$ is the model potential which has the form of eqn.~\ref{eq-bdev}. 
Equation~\ref{eq-bdev} is an inverse power law equation
$V(\phi)=V_0\phi^{-p}$  which has been previously studied by \citet{thm18} for quintessence with 
integer powers of $p$ equal to or greater than one.  In BDE the power is less than one and fractional, 
however, many of the procedures used in \citet{thm18} are also valid here. From that study it is known 
that for inverse potentials of the form $V(\phi) \propto \phi^{-p}$, where $p$ is a constant, that
\begin{equation} \label{eq-bfbde}
\beta(\phi)=\frac{p}{\kappa \phi} =\frac{\beta_b}{\kappa \phi}
\end{equation}
where $\beta_b=\frac{2}{3}$ is the BDE beta constant.  Putting eqn.~\ref{eq-bfbde} into 
eqn.~\ref{eq-beta} and integrating yields
\begin{equation} \label{eq-bdephi}
\kappa \phi=\sqrt{2 \beta_b \ln(a)+(\kappa \phi_0)^2}
\end{equation}
where $\phi_0$ is the current value of $\phi$.  Using eqn.~\ref{eq-phio}
\begin{equation} \label{eq-po}
\kappa \phi_0 = \frac{\beta_b}{\sqrt{3\Omega_{\phi_0}(w_0+1)}}
\end{equation}
where $w_0$ is the current value of the dark
energy equation of state.  Equation~\ref{eq-po} is only valid if the dark energy 
density and pressure have the quintessence forms given in eqns.~\ref{eq-rhop}.
Equations~\ref{eq-bdephi} and~\ref{eq-po} provide the means to change the analytic
parameter solutions in terms of the unobservable scalar $\phi$ into analytic
solutions in terms of the observable scale factor $a$.  Note that eqn.~\ref{eq-po}
gives a general relation between $\phi_0$ and $w_0$ but for the BDE cosmology
only one value of $\phi_0$ is allowed for a given $\Lambda_c$ as was described in 
section~\ref{ss-bder}.

\section{Calculating the Evolution of the Cosmological  Parameters and Fundamental Constants} \label{s-ev}
This section investigates whether there is a cosmological model that has the BDE potential in 
eqn.~\ref{eq-bdev}, has values of $\Lambda_c$ within $2\sigma$ of the AM19 most likely value
and has current day values of the cosmological parameters that satisfy the observational constraints.
Once $H_0$, $\Omega_{\phi_0}$ and $w_0$ are chosen $\Lambda_c$ is set by eqn.~\ref{eq-lamc}.
A single model with reasonable current values and whose past evolution is consistent with observational 
constraints is sufficient to validate the cosmology.  Here three different models are chosen where
$H_0$ and $\Omega_{\phi_0}$ are held constant but $w_0$ takes on three different values yielding
three different values of $\Lambda_c$.  This shows what parameters and evolutions are sensitive to $w_0$
and which are not.

\subsection{The model parameter space} \label{ss-mps}
The model parameters are chosen to be close to $\Lambda$CDM to explore whether BDE cosmologies
can give similar results to $\Lambda$CDM for physics quite different from $\Lambda$CDM and the 
standard model. The value of $H_0$ is set to 65 (km/sec)/Mpc and $\Omega_{\phi_0}$
is set to 0.7.  The values of $w_0$ are set to -0.99, -0.98, -0.97 all of which are well inside current
bounds on the deviation of $w$ from minus one.  These values also produce very accurate 
beta potential correspondence to the model potential in eqn.~\ref{eq-bdev}.  As shown in
fig~\ref{fig-lch} the appropriate values of $\Lambda_c$ are 53.8477, 51.2466 and 49.7837 eV 
respectively.  These values are higher that the most likely value found by AM19 of 44.02 eV or 
the theoretical value of $34_{-11}^{+16}$ eV derived by AM19.  Assuming that the quoted bounds 
on the theoretical value are $1\sigma$ even the highest value of $\Lambda_c$ (for $w_0 = -0.99$) 
is well within the $2\sigma$ limit. From eqn.~\ref{eq-ac} the scale factors $a_c$ for the 
$\Lambda_c$ values are $2.03\times10^{-6}$, $2.14\times10^{-6}$ and $2.20\times10^{-6}$.
The corresponding to condensation times are 53.0, 57.1 and 59.6 years after the big bang
about 10 years earlier than the most likely AMMA time.   

\subsection{Evolution of the Scalar} \label{ss-es}
\begin{figure}
\scalebox{.6}{\includegraphics{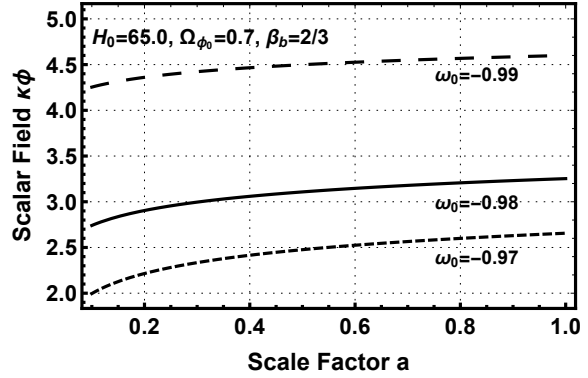}}
\caption{The evolution of the dimensionless $\kappa\phi$ for the three values of $w_0$. The 
$w_0=-0.99$ case is shown with a long dashed line, the $w_0=-0.98$ with a solid line and the 
$w_0=-0.97$ case with a short dashed line.  Unless otherwise stated this convention is used in all figures.}
\label{fig-phi}
\end{figure}

Equation~\ref{eq-bdephi} gives the evolution of the dimensionless  $\kappa\phi$ as a function 
of the scale factor $a$. This evolution is plotted in fig.~\ref{fig-phi}. 
Since $\kappa \approx 5$ $\frac{1}{M_p}$ the scalar $\phi$ is on the order of one $M_p$ or less 
for all three cases and is slowly varying.  The value of the scalar is monotonically rising with no inflections.  
This is an important input for analyzing the evolution of the dark energy equation of state $w$ at the scale 
factors considered in this work since the evolution differs from that of AMMA.

\subsubsection{Limitations on the range of the scale factor} \label{sss-leos}
Examination of eqn.~\ref{eq-bdephi} reveals, since the natural log of the scale factor is negative,
that the argument of the square root will be negative at some time in the past.  The beta function
equation for the scalar will be invalid at times earlier than this and also inaccurate near this 
region.  The scale factor where the square root argument becomes negative is determined by the
value of the current day scalar value $\phi_0$.  As is evident in fig.~\ref{fig-phi} the value of $\phi_0$
decreases with an increase in the deviation of $w_0$ from minus one.  Since the range of $w_0$ values 
considered here is very close to minus one this limitation is not a factor in a BDE cosmology at the scale 
factors and $w_0$ values used in this work.  Use of the MCMC value of $w_0=-0.9296$, however, would
approach the region of inaccuracy at small values of the scale factor.

\subsection{The Evolution of the Beta Function} \label{ss-ebf}
\begin{figure}
\scalebox{.6}{\includegraphics{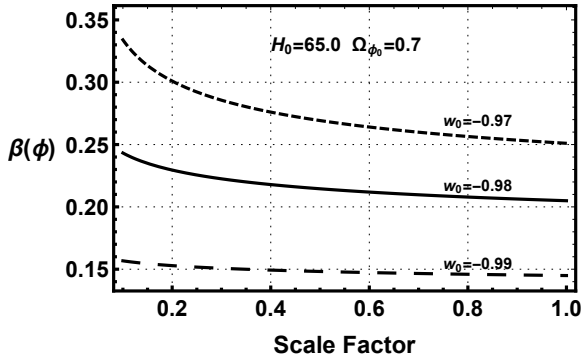}}
\caption{The evolution of $\beta(\phi)$ for the three values of $w_0$.  As defined in eqn.~\ref{eq-beta},
$\beta(\phi)$ is dimensionless.}
\label{fig-beta}
\end{figure}

Although not an observable, the evolution of the beta function is of interest due to its' central role in
the formalism. Figure~\ref{fig-beta} shows the evolution of $\beta(\phi)$ for the three values of $w_0$.
As expected from the $\kappa \phi$ plots $\beta(\phi)$ evolves more for small values of the scale factor $a$
and larger deviations of $w_0$ from minus one.  This is consistent with the general behavior of inverse
power law potential beta functions examined in \citet{thm18}.

\subsection{The Evolution of the Potential} \label{ss-epot}
The beta function methodology has two potentials, the model potential, given by eqn.~\ref{eq-mv},
which for the BDE beta function is
\begin{equation} \label{eq-mvb}
V_m=3\Omega_{\phi_0}( \frac{H_0}{\kappa})^2(\frac{\phi}{\phi_0})^{-\frac{2}{3}}
\end{equation}
The factor of $\Omega_{\phi_0}$ appears in the equation because the potential is for the bound dark 
energy only.  The second potential is the beta function potential \citep{cic17}
\begin{equation} \label{eq-vbb}
V_b=3\Omega_{\phi_0}( \frac{H_0}{\kappa})^2\exp\{-\kappa \int_{\phi_0}^{\phi}\beta(x)dx\}(1-\frac{\beta^2(\phi)}{6})
\end{equation}
yielding
\begin{equation} \label{eq-vb}
V_b(\phi)=3\Omega_{\phi_0}( \frac{H_0}{\kappa})^2(\frac{\phi}{\phi_0})^{-\frac{2}{3}}(1-\frac{\beta^2(\phi)}{6})
\end{equation}
which is the model potential multiplied by $(1-\frac{\beta^2(\phi)}{6})$.  The beta function potential
is therefore not an exact representation of the model potential but is an accurate representation as
long as $\frac{\beta^2(\phi)}{6} \ll 1$.  The accuracy of fit is examined in section~\ref{sss-af} where
it is shown that all cases fit to better than $0.8\%$ at all considered scale factors and fit to better than
$0.3\%$ for all values of $w_0$ except $w_0=-0.97$.  The accuracy of fit to the potential is representative
of the accuracy of fit to the cosmological parameters. The beta potentials match the model potentials at 
the present time which is where the boundary conditions are imposed. 

\subsubsection{Accuracy of Fit} \label{sss-af}
\begin{figure}
\scalebox{.6}{\includegraphics{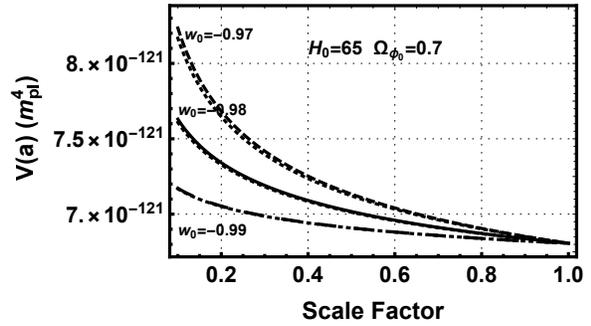}}
\caption{The evolution of model and beta potentials for the three values of $w_0$. The model potentials
are shown according to the convention of fig.~\ref{fig-phi} and the beta potentials are shown with dotted 
lines which are barely resolved from the model lines at the resolution of the figure.}
\label{fig-vf}
\end{figure}

\begin{figure}
\scalebox{.6}{\includegraphics{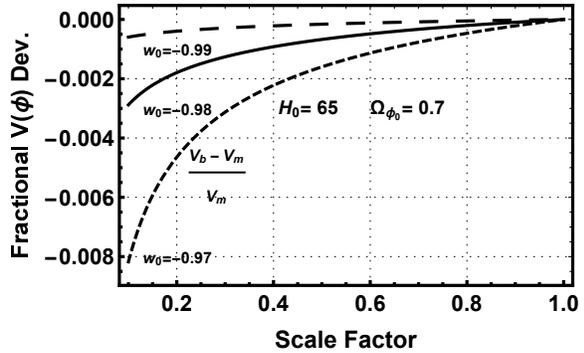}}
\caption{The fractional difference between the model and beta potentials for the three values of $w_0$.}
\label{fig-err}
\end{figure}

Figure~\ref{fig-vf} shows the model and beta potentials for the three values of $w_0$.
The potentials are decreasing with scale factor and the beta potentials lie slightly below the model
potentials but at the resolution of fig.~\ref{fig-vf} are barely resolved from the model lines.  The accuracy 
of fit is excellent and increases as the value of $w_0$ approaches minus one. A more quantitative view of 
the accuracy is provided by fig.~\ref{fig-err} where the fractional difference between the two potentials 
is plotted.
Figure~\ref{fig-err} more clearly demonstrates the improvement of the fit as $w_0$ approaches minus
one.  The maximum error is $0.8\%$ for $w_0 =-0.97$ at a scale factor of 0.1.  The errors are below
$0.3 \%$ for all of the other values of $w_0$ at all of the scale factors considered here.  The accuracy
of the potentials supports the use of the beta function formalism for this study.  The significantly improved 
accuracy of the beta function potentials over the accuracy in \citet{thm18} is because $w_0$ is very 
close to minus one for the three values considered here.

\section{Parameter Evolution with BDE and Mass} \label{s-bdem}
The beta function formalism has its roots in particle physics and string theory therefore it utilizes
a superpotential $W(\phi)$ given by
\begin{equation} \label{eq-sp}
W(\phi) = -2 H(\phi)
\end{equation}
where $H(\phi)$ is the Hubble parameter.  The superpotential plays an important role in the formalism.
To be consistent with the beta function literature $W(\phi)$ is retained as the primary
calculation tool, keeping in mind that any calculation of $W(\phi)$ is also a calculation
of $H(\phi)$.  In this work a capital $W$ always refers to the superpotential and a lower
case $w$ refers to the dark energy equation of state. 

A proper analysis of the predictions of the bound dark energy cosmology must include mass 
as was done in AMMA.  The effect of including baryonic and dark matter is most 
noticeable in the superpotential which takes on 
a new form in the presence of mass \citep{cic17}.  The first step is to calculate the evolution of the mass 
density which is independent of BDE as is shown in section~\ref{ss-md}.

\subsection{The Matter Density} \label{ss-md}
The dark energy potentials are independent of matter but both baryonic and dark matter must be 
taken into account to calculate accurate analytic solutions  for fundamental constants and cosmological 
parameters. From \citet{cic17} the matter density as a function of the scalar is given by
\begin{equation} \label{eq-rhomphi}
\rho_m(\phi)=\rho_{m_0}\exp(-3 \kappa \int_{\phi_0}^{\phi} \frac{d \phi'}{\beta(\phi')})
\end{equation}
where $\rho_{m_0}$ is the present day mass density.
Different beta functions produce different functions for $\rho_m$ as a function of $\phi$ hiding
the universality of the matter density when expressed as a function of the scale factor $a$
\begin{equation} \label{eq-rhoma}
\rho_m(a)=\rho_{m_0}\exp(-3\int_1^a d\ln(a') )= \rho_{m_0}a^{-3}
\end{equation}
as expected, independent of $\beta(\phi)$.

\subsection{The Superpotential $W$ and the Hubble Parameter $H$} \label{ss-wh}
The next step in the analysis is the calculation of the superpotential $W(\phi)$ which is also a calculation
of $H(\phi)$ by the definition of $W$ in eqn.~\ref{eq-sp}. In the presence of matter $W(\phi)$ is 
defined by a differential equation \citep{cic17}
\begin{equation} \label{eq-difw}
\frac{WW_{,\phi}}{\kappa^{3}} +\frac{\beta W^2}{2\kappa^2} = -2\frac{\rho_m}{\beta}
\end{equation}
where $W_{,\phi}$ is the derivative of $W$ with respect to $\phi$.

A general method for solving this equation is given in \citet{thm19} but here the specific method for
inverse power law potentials from \citet{thm18} is used.  The key is the use of integrating functions
to make the left side of eqn.~\ref{eq-difw} an exact integral.  The integrating function for inverse
potentials is $\phi^{\beta_b}$ where $\beta_b = \frac{2}{3}$ for BDE.  As shown in \citet{thm18}
\begin{align} \label{eq-awai}
W(a)=-\{-\frac{4\rho_{m_0}}{3\kappa^3}(\frac{2\beta_b}{3})^{-\frac{\beta_b}{2}}\exp(\frac{3(\kappa\phi_0)^2}{2\beta_b})(\kappa\phi(a))^{-\beta_b}\nonumber \\
[ \Gamma(1+\frac{\beta_b}{2},3 \ln(a)+\frac{3(\kappa\phi_0)^2}{2\beta_b}) - \Gamma(1+\frac{\beta_b}{2},\frac{3(\kappa\phi_0)^2}{2\beta_b})]\nonumber  \\
\noindent +W_0^2(\frac{\phi_0}{\phi(a)})^{\beta_b}\}^{1/2}
\end{align}
where $\Gamma$ is the incomplete Gamma function and $\kappa\phi(a)$ is given by eqn.~\ref{eq-bdephi}.  
Substituting the BDE $\beta_b=\frac{2}{3}$ gives
\begin{align} \label{eq-awai}
W(a)=-\{-\frac{4\rho_{m_0}}{3\kappa^3}(\frac{4}{9})^{-\frac{1}{3}}\exp(\frac{9(\kappa\phi_0)^2}{4})(\kappa\phi(a))^{-\frac{2}{3}}\nonumber \\
[ \Gamma(\frac{4}{3},3 \ln(a)+\frac{9(\kappa\phi_0)^2}{4}) - \Gamma(\frac{4}{3},\frac{9(\kappa\phi_0)^2}{4})]\nonumber  \\
\noindent +W_0^2(\frac{\phi_0}{\phi(a)})^{\frac{2}{3}}\}^{1/2}
\end{align}
Of course $W_0$ is just $-2H_0$.

\subsubsection{The Hubble Parameter as a Function of the Scale Factor} \label{sss-eh}
\begin{figure}
\scalebox{.6}{\includegraphics{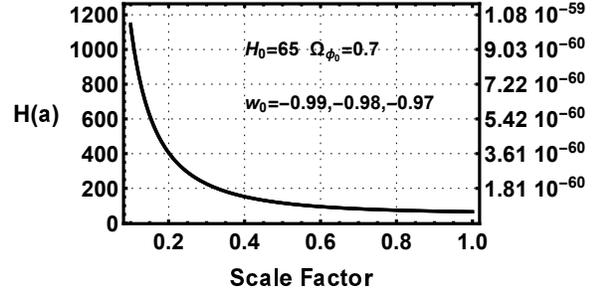}}
\caption{The evolution of the Hubble parameter in (km/sec)/Mpc on the left and in $M_p$ on the right 
as a function of the scale factor $a$ for the three  values of $w_0$.  The three plots are identical to the 
width of the line.}
\label{fig-hplt}
\end{figure}

Figure~\ref{fig-hplt} shows the evolution of the Hubble parameter in the familiar units of
(km/sec)/Mpc on the left ordinate and in $M_p$ on the right ordinate.  Note that the Hubble parameter
for all three values of $w_0$ are plotted in fig.~\ref{fig-hplt}. As in \citet{thm18,thm19}, they are 
identical to the width of the line. This is a common feature of the Quintessence Model cosmologies 
when mass is included.

\begin{figure}
\scalebox{.6}{\includegraphics{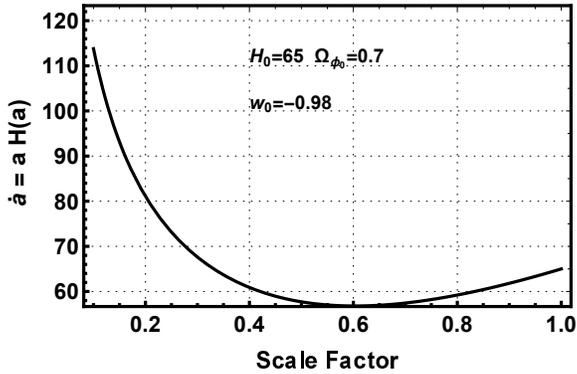}}
\caption{The evolution of $\dot{a}$ with respect to the scale factor $a$.}
\label{fig-adot}
\end{figure}

A first test of the BDE superpotential is whether the late time acceleration of the expansion of
the universe occurs at the correct time.  The time derivative of scale factor $\dot{a}$ is used as the 
check via
\begin{equation} \label{eq-adot}
\dot{a} = a H = -\frac{1}{2} a W
\end{equation}
Figure~\ref{fig-adot} shows the evolution of $\dot{a}$ with respect to $a$.
As expected, $\dot{a}$ decreases at early matter dominated times and then begins to 
increase at a scale factor of $\sim0.6$ consistent with the observed beginning of the dark energy 
dominated epoch.  The BDE cosmology therefore predicts an onset of late time acceleration of the
expansion of the universe that is consistent with observations.

\subsubsection{Comparison of the Hubble Parameter with observations} \label{sss-ho}
\begin{figure}
\scalebox{.6}{\includegraphics{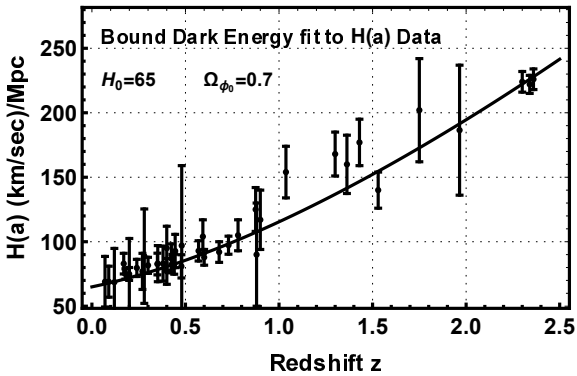}}
\caption{A comparison of the BDE $H$ evolution, solid line, to the observational compilation
of $H$ values by \citet{jes17}, data points with errors.}
\label{fig-hfit}
\end{figure}

A second test of the BDE superpotential is whether the Hubble Parameter is consistent with the
current observations.  The test is performed with a recent compilation of Hubble parameter
measurements in \cite{jes17} as was done in \citet{thm18,thm19}.  These measurements
are taken simply as a typical compilation with no judgement as to their quality relative to
other compilations.  The results are shown in fig.~\ref{fig-hfit}.  It is interesting to note that
a Chi square fit to the data in \citet{thm18} found a best fit value of $H_0$ of 66.5 rather
close to the value of 65 used here.  As a result the fit shown in fig.~\ref{fig-hfit} is slightly better
than in \citet{thm18} which used a fiducial value of $H_0=70$ for all of the examined potentials.
It is obvious that the BDE evolution of $H$ is consistent with the observational data.  It should
be noted that \citet{thm18,thm19} showed that the evolution of $H$ is very insensitive to the 
form of the dark energy potential so that the slightly better fit to the observations is due to the
change in $H_0$ and not due to the different BDE potential.  It should be further noted that to the
thickness of the plotted line all three $w_0$ cases are indistinguishable. 

\subsubsection{Comparison with $\Lambda$CDM}
\begin{figure}
\scalebox{.6}{\includegraphics{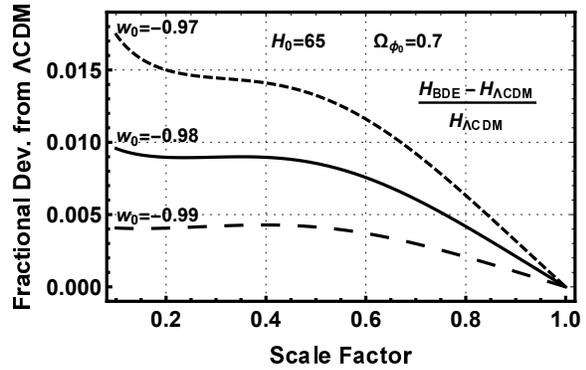}}
\caption{The figure indicates the fractional deviation of the BDE $H(a)$ from the $\Lambda$CDM
$H(a)$ with $H_0$ set to 65 for both cosmologies.}
\label{fig-hdev}
\end{figure}

A primary goal of current cosmology is to determine whether dark energy is static, $\Lambda$CDM
or dynamic, eg. rolling scalar fields.  It is already known \citep{thm18,thm19} that for quintessence
cosmologies with various forms of the dark energy potential the evolution of the Hubble parameter 
$H(a)$ is virtually indistinguishable from $\Lambda$CDM.  Since the BDE cosmology has the
quintessence forms of the dark energy pressure and density it is expected that it too will be close
to the $\Lambda$CDM evolution.  Figure~\ref{fig-hdev}, which plots $(H(a)_{BDE}-H(a)_{\Lambda CDM})
/H(a)_{\Lambda CDM}$, shows that this is the case.  It also shows that the BDE Hubble parameter
is slightly larger at earlier times than the $\Lambda$CDM Hubble parameter for the scale factors considered 
in this work where $H_0 = 65$ for both cases.
The maximum deviation from $\Lambda$CDM for the $w_0=-0.97$ case is only $1.8\%$ at a
redshift of 9 and below $1.0\%$ for the other $w_0$ values and at lower redshifts.
This indicates that it is very hard to distinguish the dynamic BDE cosmology from the static
$\Lambda$CDM cosmology based only on the evolution of the Hubble parameter.  An 
interesting aspect of fig.~\ref{fig-hdev} is that most of the evolution of BDE away from
$\Lambda$CDM occurs at redshifts between 0 and $\sim1$ with relatively little evolution at 
higher redshifts.  The deviation from $\Lambda$CDM at small scale factors is due to the identical
values of $H_0$ for both cases even though both are evolving in a matter dominated epoch at
small scale factors.  Comparison with figure 8 of AM19 shows that if BDE curve was raised to zero 
$\Delta\Lambda$CDM at zero redshift there would be the same offset.

\subsection{The Dark Energy Density $\Omega_{\phi}$} \label{ss-omp}
\begin{figure}
\scalebox{.6}{\includegraphics{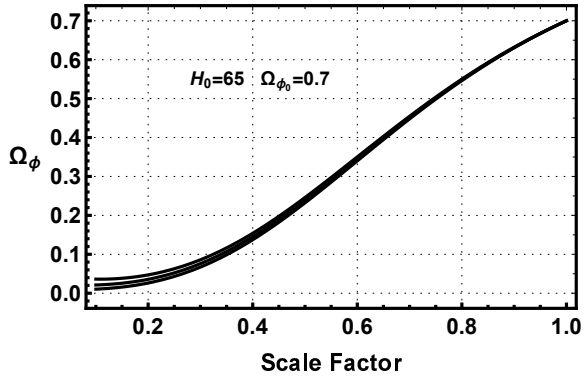}}
\caption{The evolution of  $\Omega_{\phi}$ with  $\Omega_{\phi_0}=0.7.$ At the resolution
of the plot the lines for the three values of $w_0$ overlap}
\label{fig-omp}
\end{figure}

The evolution of the dark energy density with respect to the critical density $\Omega_{\phi}$ is an 
observable parameter of interest.  In a flat cosmology $\Omega_{\phi}$ is the difference between
the total density and the mass density over the total density
\begin{equation} \label{eq-omp}
\Omega_{\phi}=\frac{3(H/\kappa)^2-\rho_{m_0}a^{-3}}{3(H/\kappa)^2}
\end{equation}
Figure~\ref{fig-omp} shows the evolution of $\Omega_{\phi}$ for the BDE cosmology with mass.
It is clear that the evolution of  $\Omega_{\phi}$ is not a strong function of $w_0$ but the
values of $\Omega_{\phi}(a)$ for the three $w_0$ at $a=0.1$ are not exactly equal as is
indicated by the slight separation of the lines in fig.~\ref{fig-omp} as they approach $a=0.1$.

\subsection{The Time Derivative of the Scalar} \label{ss-pdot}
\begin{figure}
\scalebox{.6}{\includegraphics{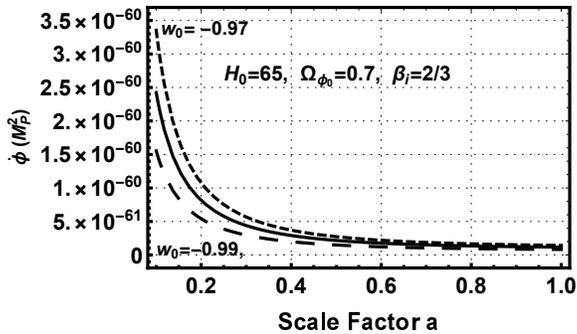}}
\caption{The evolution of $\dot{\phi}$ as a function of the scale factor $a$.  The figure
uses the same line style convention as previous figures for the three values of $w_0$}
\label{fig-pdot}
\end{figure}

The time derivative of the scalar $\dot{\phi}$ is not an observable but is an important
parameter as demonstrated by eqns.~\ref{eq-rhop}.  A quick examination of the 
definitions of $\beta$ and $H$ shows that
\begin{equation} \label{eq-pdot}
\kappa \dot{\phi} =\beta H
\end{equation}
Figure~\ref{fig-pdot} shows the evolution of $\dot{\phi}$ as a function of the scale factor.
The magnitude of $\dot{\phi}$ is decreasing monotonically in time with no inflection points.
This is consistent with the evolution of $\phi$ shown in fig.~\ref{fig-phi}.

\subsection{The Dark Energy Equation of State} \label{ss-bdes}
\begin{figure}
\scalebox{.6}{\includegraphics{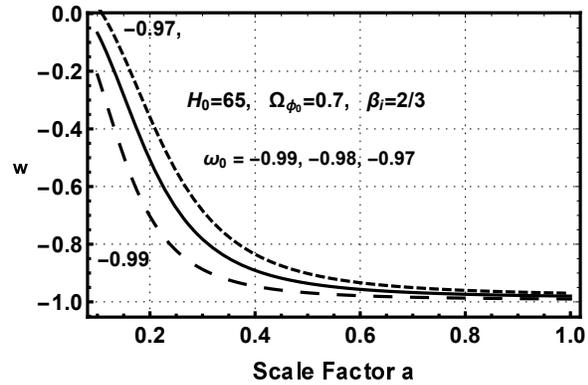}}
\caption{The evolution of the BDE equation of state $w$ with respect to the scale factor for
the three initial values of $w$. with the standard line style conventions for the values of $w_0$.}
\label{fig-w}
\end{figure}

From \citet{cic17} and \citet{thm19} the dark energy equation of state in a cosmology with the
quintessence density and pressure relations is
\begin{equation} \label{eq-wden}
1+w(\phi) = \frac{\beta^2}{3}\frac{1}{(1-\frac{4\rho_{m_0}a^{-3}}{3(W/\kappa)^2})}
=\frac{\beta^2}{3}\frac{1}{(1-\Omega_m)}=\frac{\beta^2(\phi)}{3\Omega_{\phi}}
\end{equation}
This differs slightly from \citet{thm19} in that the $\kappa$ factors are included in the definition
of the beta function given by eqn.~\ref{eq-beta}.  Using the BDE $\beta(\phi)$ fig.~\ref{fig-w} 
shows the evolution of $w$ for the three values of $w_0$.
Figure~\ref{fig-w} shows the classic freezing evolution of $w(a)$ which monotonically decreases
toward minus one.  This is consistent with the general inverse power law behavior studied in
\citet{thm18} but different from the evolution in AMMA which shows an inflection near $z=3$
to an increasing $w$ rising to -0.93 at $z=0$.  No such inflection is found in this work or in 
\citet{thm18, thm19} for any of the studied potentials.

The $w$ evolution has some interesting features.  The values of $w$ stay close to their associated
$w_0$ for scale factors between the present day value of one and a value of ~0.5 which is roughly
half the age of the universe.  Since the three $w_0$ values are very close to minus one this is not
surprising and follows the general pattern of inversed power potentials found in \citet{thm18}. It
is therefore difficult to detect a dynamical universe with observations at redshifts less than one.  At
smaller scale factors, redshifts greater than one, there is significant evolution of $w$ that should be
detectable.

\section{The Fundamental Constants} \label{s-fc}
A powerful, but seldom used, cosmological parameter is the measurement of fundamental constants
in the early universe.  Here fundamental constants are restricted to dimensionless constants such as
the fine structure constant $\alpha$ and the proton to electron mass ratio $\mu$.  If the scalar field
responsible for the acceleration of the expansion of the universe interacts with sectors other than
gravity it can alter the values of the fundamental constants.  AMMA states that BDE interacts
only with gravity.  If this is the case it should have no effect on the fundamental constants.  However,
without evoking finely tuned symmetries it is very difficult to create a scalar field that only interacts with
gravity and does not have couplings to any other sector \citep{car98}.  It is therefore useful to determine
the limits on the coupling to other sectors imposed by the limits on the variation of $\alpha$ and $\mu$.
If the BDE scalar field couples with $\alpha$ and $\mu$ there
is a relationship between the variance of $w$ and the variance of the fundamental constants
\citep{cal11,thm12}.  In this work the proton to electron mass ratio $\mu$ is used as an example since
it has more stringent limits on its variation than the current limits on $\alpha$.  The limit on the variation
of $\mu$ is $\Delta \mu / \mu \le 10^{-7}$ at a redshift of 0.892 \citep{bag13,kan15} which is a lookback
time of a little over half the age of the universe.

\subsection{The Stability of $\mu$}\label{ss-dmu}
The relationship between the variation of $\mu$ or $\alpha$ and $\phi$ is given simply by
\begin{equation} \label{eq-dx}
\frac{\Delta x}{x} = \zeta_x \kappa(\phi-\phi_0)
\end{equation}
where $x$ is either $\mu$ or $\alpha$ and $\zeta_x$ is the dimensionless coupling constant for the interaction. 
This can be thought of as the first term in a Taylor series expansion of a more complicated coupling where
the second term would be on the order of $10^{-7}$ smaller than the first.  The coupling is actually
a mixture of couplings with the Quantum Chromodynamic Scale, the Higgs Vacuum Expectation Value and
the Yukawa couplings \citep{coc07,thm17}.

\begin{figure}
\scalebox{.6}{\includegraphics{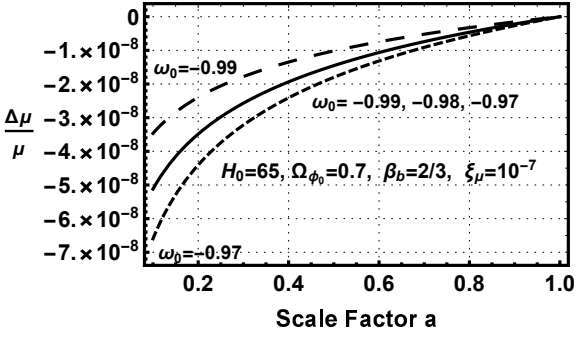}}
\caption{The evolution of the $\frac{\Delta \mu}{\mu}$ respect to the scale factor for
the three initial values of $w$. with the standard line style conventions for the values of $w_0$.}
\label{fig-dmu}
\end{figure}

Figure~\ref{fig-dmu} shows the evolution of $\Delta \mu / \mu$ for a coupling constant of $\zeta_{\mu}
=  10^{-7}$ which is conservatively small.
All of the tracks satisfy the observational constraint on the variation of $\mu$.  Since $\phi$ is
increasing with time $\phi_0$ is always larger than $\phi$, therefore, the variation of $\mu$ is 
negative for a positive $\zeta_{\mu}$, however, the coupling could equally well be negative 
resulting in a positive $\Delta \mu$.  From eqn.~\ref{eq-dx} it is clear that the calculation of
the evolution of $\alpha$ is identical except for the value of the coupling constant $\zeta_{\alpha}$.

\subsection{The Relation between $\Delta \mu/\mu$ and $w$} \label{ss-muw}
\begin{figure}
\scalebox{.6}{\includegraphics{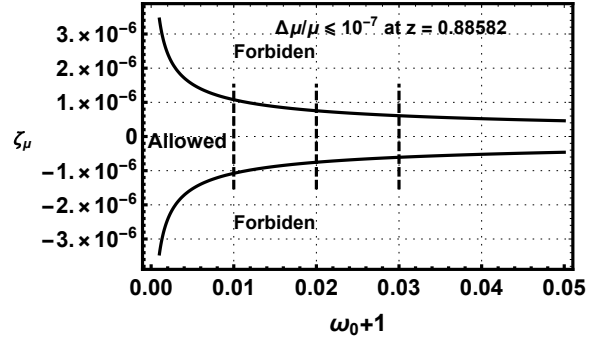}}
\caption{The allowed and forbidden regions in the $\zeta_{\mu}$ $(w_0+1)$ plane imposed by
the observed limits on $\Delta \mu / \mu$.  The three dashed lines indicate the values of $(w_0+1)$
used here.}
\label{fig-zw}
\end{figure}

The observational restrictions on the evolution of either $\mu$ or $\alpha$ can be met by either
lowering the value of the coupling or by lowering the value of $(w_0+1)$.  The explicit relation
between $w_0$, $\Delta \mu/\mu$ and $\zeta_{\mu}$ from eqns.~\ref{eq-bdephi},~\ref{eq-phio}
and~\ref{eq-dx}
\begin{equation} \label{eq-uz}
\zeta_{\mu} =\frac{\Delta \mu / \mu}{\sqrt{2\beta_b\ln(a_{ob})+\frac{\beta_b^2}{3\Omega_{\phi_0}(w_0+1)}}-\frac{\beta_b}{\sqrt{3\Omega_{\phi_0}(w_0+1)}}}
\end{equation}
where $a_{ob}$ is the scale factor of the observation of the limiting constraint and again for BDE,
$\beta_b=\frac{2}{3}$. Figure~\ref{fig-phi}
demonstrates that the closer $w_0$ is to minus one the lower the change is in $\phi$ for a given 
$\Delta a$. Equations~\ref{eq-bdephi} and~\ref{eq-po} show the relationship between the
evolution of $\phi$ and $w_0$.  Observational constraints on $\Delta \mu / \mu$ then define
allowed and forbidden areas in the $\zeta_{\mu}$ $(w_0+1)$ plane as shown in figure~\ref{fig-zw}.
$\Lambda$CDM and the standard model occupy the (0.0,0.0) position in fig.~\ref{fig-zw}. The three
dashed lines indicate the values of $(w_0+1)$ used in this work.  For $w_0 = -0.99$ fig.~\ref{fig-zw}
indicates that $\zeta_{\mu}$ must lie between $\pm 10^{-6}$ while for $w_0=-0.97$ the allowed
value of $\zeta_{\mu}$ is reduced to $\pm 6 \times 10^{-7}$.  If BDE only interacts with gravity 
then this limit is satisfied by definition. If BDE actually does couple to the other sectors then these 
limits impose constraints on the couplings that are sensitive to the value of $w_0$.
                                                                                                                                                                                                
\section{Checking on the Swampland} \label{ss-cs}
There is currently vigorous discussion of the swampland conjectures \citep{vaf05,agr18} that
defines a parameter space, the swampland, that is incompatible with a quantum theory
of gravity.  Avoiding the swampland requires that both the change in the scalar should be
$\Delta \phi < \sim O(1)$ in reduced Planck mass units and that $|\frac{dV}{d\phi V}| \
\geq \sim O(1)$.  As discussed in \citet{thm19}
the inverse power law potentials generally comply with the first criteria as is the case here. 
Remembering that in our reduced Planck mass units that $\kappa=1$ figure~\ref{fig-phi} shows 
that in the range considered here $\Delta \phi$ is less than 0.5 for all  cases and is slowly increasing 
with time making $\Delta \phi$ small.  For the second criteria  $\frac{dV}{d\phi V}=-\frac{2}{3\phi}$ 
making the largest value 1/3 for $w_0=-0.97$ at a scale factor of 0.1 and 
smaller than 1/3 for the other cases.  Depending on the interpretation of greater than order one
it might be argued that the $w_0=-0.97$ and -0.98 cases are on the fringe of compliance but
the $w_0=-0.99$ case is clearly not.  

\section{Summary} \label{s-sum}
The results of this investigation show that the BDE cosmology is compatible with the
observational constraints on the current value and evolutional history of the
Hubble parameter, $H$, the matter density $\rho_m$, the dark energy equation of
state, $w$ and the fundamental constants $\mu$ and $\alpha$. It is compatible with
the swampland criteria on the evolution of the scalar $\phi$ but does not strictly satisfy the
criterion on the evolution of the potential $V(\phi)$ as is typical of power and inverse
power law cosmologies with quintessence pressure and density forms.  

The investigation also produced useful relations between $H$, the condensation scale
$\Lambda_c$, $\Omega_{\phi}$
and $w$.  These relations, coupled with the beta function formalism, enabled the accurate 
analytic calculation of the evolution of the cosmological parameters and fundamental 
constants for scale factors between 0.1 and 1.0.  In the calculations the value of the
Hubble constant was fixed at 65 (km/sec)/Mpc and the current value of the ratio of 
the dark energy density, $\Omega_{\phi_0}$ was set to 0.7.  Three different values
of $w_0$ close to minus one were used.  Since the BDE cosmology is prescriptive
each $w_0$ value requires a different $\Lambda_c$ calculated using eqn.~\ref{eq-lamc}.
The calculated $\Lambda_c$ values are higher than the most likely value found by
AMMA but at a less than $2\sigma$ difference.  The inflection of w from decreasing
to increasing at $z=3$ found by AMMA is not confirmed in this study.  It appears that
the BDE cosmology passes the cosmological and fundamental constant constraints
considered here.  Tests of the validity of the particle physics derivation of BDE is
beyond the scope of this work.

\label{lastpage}
\end{document}